\documentclass[letterpaper]{article} 
\usepackage{amsfonts}
\usepackage{aaai2026}  
\usepackage{times}  
\usepackage{helvet}  
\usepackage{courier}  
\usepackage[hyphens]{url}  
\usepackage{graphicx} 
\usepackage{natbib}  
\usepackage{caption} 
\usepackage{algorithm}

\usepackage{newfloat}
\usepackage{listings}
\usepackage{amsmath,amssymb,amsfonts}
\usepackage{graphicx}
\usepackage{textcomp}
\usepackage{amsmath}
\usepackage{cases}
\usepackage{diagbox}
\usepackage{booktabs}
\usepackage{graphicx}
\usepackage{subfigure}
\usepackage{algpseudocode}
\usepackage{amsmath}
\usepackage{graphics}
\usepackage{epstopdf}
\usepackage{bm}
\usepackage{amssymb,amsfonts}
\usepackage{color}
\usepackage{float}
\usepackage{lineno}

\usepackage{url,multirow}

\usepackage{newfloat}
\usepackage{listings}
\DeclareCaptionStyle{ruled}{labelfont=normalfont,labelsep=colon,strut=off} 
\floatstyle{ruled}
\newfloat{listing}{tb}{lst}{}
\floatname{listing}{Listing}
\title{ASLSL: Adaptive shared latent structure learning with incomplete multi-modal physiological data for multi-dimensional emotional feature selection}
\author{
    \textbf{Xueyuan Xu*}, \textbf{Tianze Yu}, \textbf{Wenjia Dong}, \textbf{Fulin Wei}, \textbf{Li Zhuo} \\
    {\normalfont School of Information Science and Technology, Beijing University of Technology, Beijing 100124, China} \\
    {\normalfont School of Artificial Intelligence, Anhui University, Beijing 100124, China} \\
    {\normalfont \{xxy, zhuoli\}@bjut.edu.cn, \{23027418, 23027425\}@emails.bjut.edu.cn,weifulin@ahu.edu.cn}
}

\usepackage{bibentry}

\begin{document}

\maketitle

\begin{abstract}
Recently, multi-modal physiological signals based emotion recognition has garnered increasing attention in the field of brain-computer interfaces. Nevertheness, the associated multi-modal physiological features are often high-dimensional and inevitably include irrelevant, redundant, and noisy representation, which can easily lead to overfitting, poor performance, and high computational complexity in emotion classifiers. Feature selection has been widely applied to address these challenges. However, previous studies generally assumed that multi-modal physiological data are complete, whereas in reality, the data are often incomplete due to the openness of the acquisition and operational environment. For example, a part of samples are available in several modalities but not in others. To address this issue, we propose a novel method for incomplete multi-modal physiological signal feature selection called adaptive shared latent structure learning (ASLSL). Based on the property that similar features share similar emotional labels, ASLSL employs adaptive shared latent structure learning to explore a common latent space shared for incomplete multi-modal physiological signals and multi-dimensional emotional labels, thereby mitigating the impact of missing information and mining consensus information. Two most popular multi-modal physiological emotion datasets (DEAP and DREAMER) with multi-dimensional emotional labels were utilized to compare the performance between compare ASLSL and seventeen feature selection methods. Comprehensive experimental results on these datasets demonstrate the effectiveness of ASLSL.
\end{abstract}


\section{Introduction}
Existing affective computing researches predominantly utilize discrete or multi-dimensional emotion models to represent distinct emotional states. Compared to discrete emotion models, multi-dimensional emotional label models offer broader characterization and are capable of describing the variations in emotional states \cite{khare2023emotion}. Currently, emotion recognition approaches predominantly encompass facial expressions, speech, body gestures, text, and neurophysiological signals\cite{ezzameli2023emotion}. Emotions arise alongside physiological and psychological activities, and neurophysiological signals offer the advantage of being difficult to fake compared to other approaches\cite{wu2023affective}. 

\begin{figure}[!t]
\centering
\includegraphics[width=0.478\textwidth]{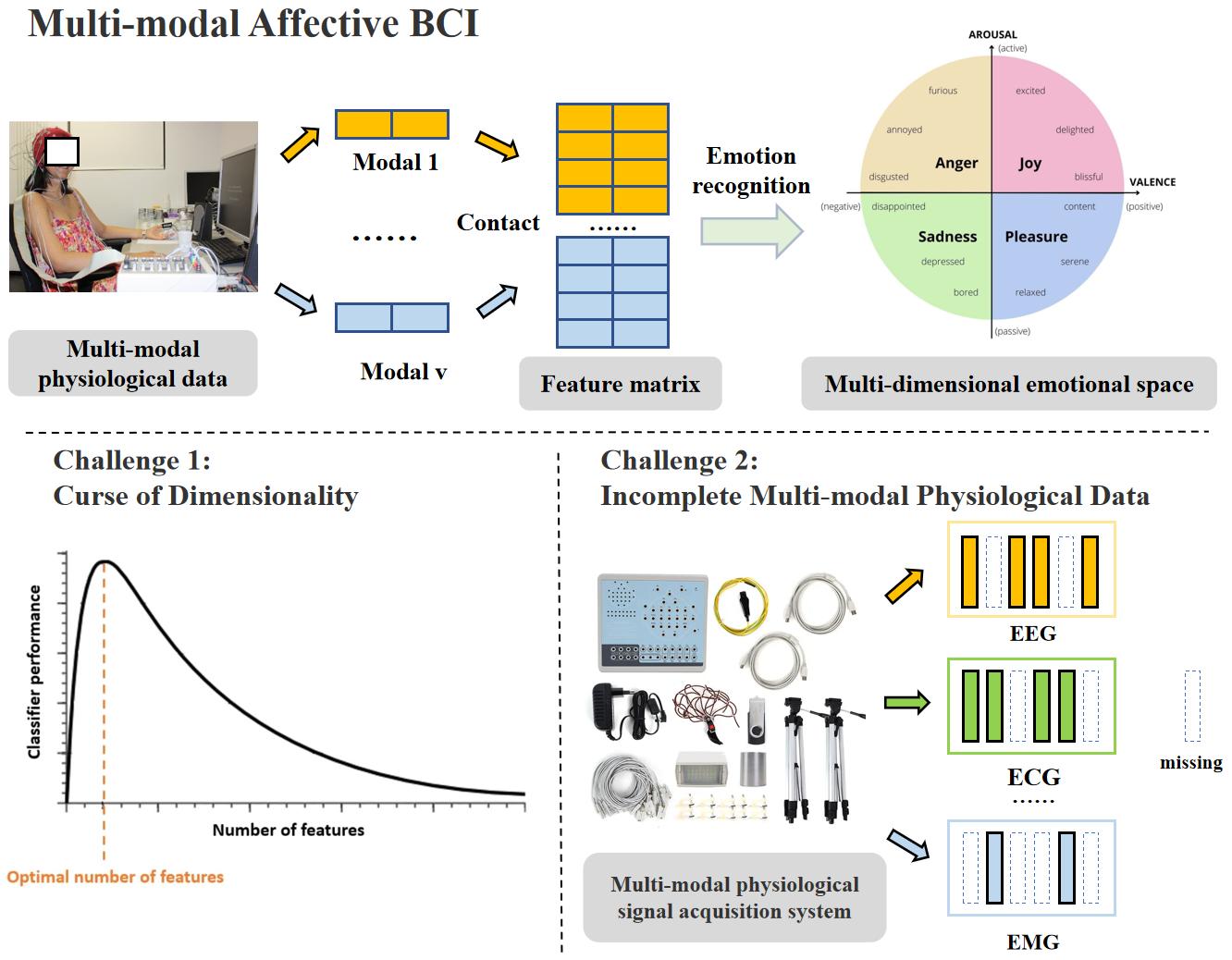}
\caption{An illustration of the multi-modal affective BCI applications, along with two challenges: (1) Incomplete multi-modal physiological signal data induced by the openness of the acquisition and operational environment, such as channel noise or electrode detachment; (2) The high dimensionality of multi-modal physiological features and the relatively small number of multi-modal samples collected synchronously cause the curse problem of dimensionality.}
\label{Illustration}
\end{figure}

Among multiple physiological signals, electroencephalogram (EEG) has garnered considerable attention in emotion recognition research due to its high temporal resolution, non-invasiveness, and ease of implementation\cite{wang2024research}. Nevertheless, due to the nonlinear, non-stationary nature and individual variability of EEG signals\cite{li2022eeg}, emotion recognition systems based on unimodal EEG often face challenges related to low accuracy in practical real-world applications\cite{geetha2024multimodal}. To address these issues, several studies have attempted to analyze emotion states using multi-modal neurophysiological signals\cite{zhang2020emotion}. Each modality reveals specific aspects of the processes and dynamics of emotions, and complementary multi-modal neurophysiological signals provide multi-level emotional representations that enhance the precision of emotion recognition\cite{geetha2024multimodal}.

The proliferation of sensors for affective computing has accelerated due to advancements in multi-modal physiological signal acquisition technologies, resulting in a substantial volume of features extracted from these sensors \cite{Becker2020HRE}. However, the limited availability of high-quality multi-modal physiological signal samples means that the resultant features are often high-dimensional and may contain redundant, irrelevant, or noisy data\cite{GRMOR2021taffc}. This can lead to challenges such as overfitting, diminished performance, and increased computational complexity \cite{wang2020emotion}. Feature selection has emerged as a crucial technique for identifying and retaining significant features while removing irrelevant and noisy ones from the data. This approach helps maintain the core physiological and psychological representations of physiological signal features, thereby enhancing the transparency and interpretability of affective computing models \cite{jenke2014taffc,liu2018electroencephalogram}.

Feature selection methods can be broadly categorized into three main types based on their feature evaluation and search mechanisms: filter methods, wrapper methods, and embedded methods \cite{li2017feature}. Filter methods assess the importance of neurophysiological signal features in emotion recognition using statistical metrics or information entropy measures. However, these methods often yield suboptimal feature selection results, regardless of the performance of classifiers\cite{zhang2019review}. To address these limitations, several studies have investigated wrapper methods, which generally offer superior classification performance compared to filter methods. This is because wrapper methods evaluate feature subsets based on the performance of a specific classifier \cite{hou2011JELSR}. Nonetheless, wrapper methods can be computationally intensive due to the extensive number of trials required \cite{zhang2019review}. Recently, embedded methods have gained attention as a viable alternative to overcome the drawbacks of filter methods. These methods integrate the process of feature selection directly into the model optimization framework. Experimental results in \cite{torres2017svm,GRMOR2021taffc} shows the effectiveness of the embedded approaches in emotion recognition.

Existing studies on physiological feature selection typically assume that multi-modal physiological signal data are complete. However, due to the inherent variability of the acquisition environment, practical applications of emotion recognition based on multi-modal neurophysiological signals often encounter incomplete sample data. For example, some samples may be available in certain modalities but not in others. This lack of complete information directly impedes the accurate modeling of the relationship between multi-modal neurophysiological signals and multi-dimensional emotional labels\cite{wang2024incomplete}.

To address this issue, as illustrated in Fig.\ref{Illustration}, we propose a novel feature selection model for incomplete multi-modal physiological feature selection, termed adaptive shared latent structure learning (ASLSL). This model utilizes adaptive shared latent structure learning to explore a common latent space for multi-modal physiological signals and multi-dimensional emotions, even in the presence of incomplete multi-modal information. Additionally, it employs adaptive learning of modality weights to assess the significance of different modalities in emotion recognition. By the above means, ASLSL could effectively uncovers the relationships between incomplete multi-modal neurophysiological data and multi-dimensional emotional representation, thereby facilitating the construction of informative emotional feature subsets.

Furthermore, the contributions of our work are as follows:

\begin{enumerate}
  \item[$\bullet$] We propose a novel incomplete multi-modal physiological feature selection method for the multi-dimensional affective computing task. ASLSL employs adaptive shared latent structure learning to explore a latent space shared for incomplete multi-modal signal data and multi-dimensional emotional representation, thereby mitigating the impact of missing information and mining consensus information. With the above strategy, ASLSL could adaptively adjust the discriminative analysis results for various feature types within the incomplete multi-modal data, thus realizing efficient and accurate multi-dimensional affective computing. To the best of our knowledge, this is the first study to construct an incomplete multi-view feature selection framework in a complete multi-label data scenario. 
  \item[$\bullet$] To address the optimization challenges associated with ASLSL, an efficient alternative approach is proposed to ensure convergence and attain an optimal solution.
  \item[$\bullet$] To validate the efficacy of ASLSL, we utilized two publicly available datasets: DREAMER and DEAP. Experimental results show that ASLSL outperforms seventeen advanced feature selection methods, achieving superior emotion recognition performance across six metrics.
\end{enumerate}

\begin{figure*}[!t]
\centering
\includegraphics[width=0.96\textwidth]{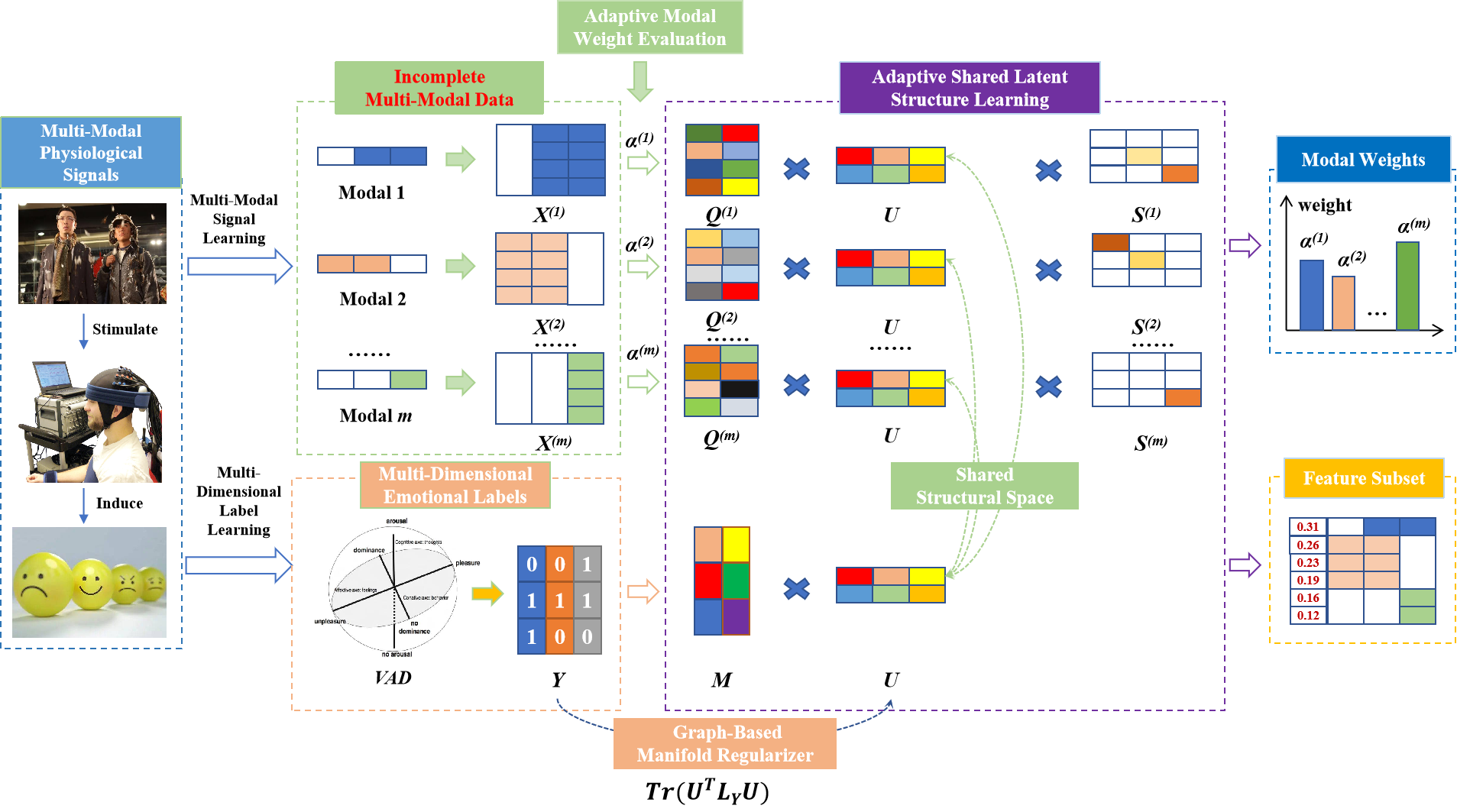}
\caption{The proposed ASLSL framework includes the following processes: (a) adaptive shared latent structure learning; (b) adaptive modal weight evaluation; (c) graph-based manifold regularizer.}
\label{Framework_ASLSL}
\end{figure*}

\section{Problem formulation}\label{Pf}
The ASLSL framework is formulated as follows:
\begin{equation}
\min _{Q^{(v)}, M, U} F(X^{(v)}, Q^{(v)}, S^{(v)}, M, U, Y)+\lambda C(U)+\gamma \Omega(Q^{(v)})
\end{equation}
where $\gamma$ and $\lambda$ are both regularization parameters.  $X = \left \lbrace X^{\left( 1 \right)}, X^{\left( 2 \right)}, \cdots , X^{\left( v \right)}, \cdots , X^{\left( m \right)} \right \rbrace$ is multi-modal physiological signal feature data, and each element $X^{\left( v \right)} =[x_1, x_2, ... , x_{d_{v}}]^T$. $X^{\left( v \right)} \in \mathbb{R}^{d_{v} \times n}$. $Y \in\{0,1\}^{k\times n}$ is a multi-dimensional emotional label matrix. $S = \left \lbrace S^{\left( 1 \right)}, S^{\left( 2 \right)}, ... , S^{\left( v \right)}, ... , S^{\left( m \right)} \right \rbrace$ where $S^{\left( v \right)} \in \mathbb{R}^{n \times n}$ is an indicator matrix that marks the location of the missing instances. $U\in \mathbb{R}^{n \times k}$, $Q^{(v)}\in \mathbb{R}^{d_{v} \times k}$, and $M\in \mathbb{R}^{k \times k}$ are shared space of all modalities, projection matrix of $X^{(v)}$, and coefficient matrix of $Y$, respectively. $d^{\left( v \right)}$, $n$, $m$, and $k$ correspond to the number of features of each modal, instances, modalities, and dimensions of labeling, respectively.

The adaptive shared latent structure term, graph-based manifold regularization term, and $l_{2,1}$-norm sparsity restriction term are represented by the symbols $F$, $C$, and $\Omega$, respectively. The subsections will introduce the definitions of $F$, $C$, and $\Omega$.

\subsection{Adaptive shared latent structure term}
Shared latent structure learning can be obtained as follows:
\begin{equation}
\begin{gathered}
\label{equ_framework_lls}
\min _{Q, M, U} \left\|X-Q U^{\mathrm{T}}\right\|_F^2+\lambda\left\|Y-M U^{\mathrm{T}}\right\|_F^2 \\
\text { s.t. } Q \geq 0, M \geq 0, U \geq 0 \\
\end{gathered}
\end{equation}
where the feature matrix $X$ can be represented by two non-negative matrices $U$ and $Q$. Similarly, the multi-dimensional emotional label matrix $Y$ can be decomposed to a latent label structure matrix $U$ and corresponding coefficient matrix $M$. Based on the property that similar features share similar emotional labels, a common matrix $U$ is utilized as the latent structure space for $X$ and $Y$, which could share dependence between physiological signal features and multi-dimensional emotional labels. 

Inspired by the previous research findings that every modal have different contributions to the emotion recognition task\cite{koelstra2011deap,zhang2020emotion}, an adaptive modal weight evaluation strategy is adopted to assign the modal-weight $\alpha^{(v)}$ ($v=1, \ldots, m$) for each physiological signal modal. Additionally, the same latent feature structure $U$ is shared by each physiological signal to mine the consensus information across various modalities. Finally, to represent the instance-missing information, a diagonal weighted matrix $S^{(v)}$ in the $v$-th modal is included. 
In conclusion, $F$ can be formulated as follows:
\begin{equation}
\begin{aligned}
\label{equ_framework_wlls}
&\min_{Q^{(v)}, \alpha^{(v)}, M, U} \sum_{v = 1}^{m} \left(\alpha^{(v)}\right)^{\gamma} \left[ \left\|\left(X^{(v)} - Q^{(v)} U^T\right) S^{(v)}\right\|_F^2  \right. \\
& \left. + \lambda \left\|Y - M U^T\right\|_F^2 \right] \\
&\text{s.t.} M \geq 0, \quad Q^{(v)} \geq 0, \quad U \geq 0, \\
& 0 \leq \alpha^{(v)} \leq 1, \quad \sum_{v = 1}^{m} \alpha^{(v)} = 1, \quad v = 1, \ldots, m
\end{aligned}
\end{equation}
where $\gamma$ and $\lambda$ are both regularization parameters.

The diagonal element $S^{(v)}_{j,j}$ of the diagonal weighted matrix $S^{(v)}$ in the $v$-th modal is defined as follows:
\begin{equation}
S^{(v)}_{j,j}= \begin{cases}1 & \text { if } j \text {-th instance exists in } v \text {-th modal; } \\ 0 & \text { otherwise. }\end{cases}
\end{equation}

\subsection{Graph-based manifold regularization term}
Based on the spectral graph theory\cite{jian2016multi}, a graph-based manifold regularizer $\operatorname{Tr}\left(U^T L_Y U\right)$ is employed to maintain consistency of the local geometric structures between the shared latent structure space $U$ and the multi-dimensional emotional label space $Y$. The graph-based manifold regularization term $C$ can be formulated as follows:
\begin{equation}
\begin{gathered}
\label{equ_framework_C}
C(U) = \operatorname{Tr}\left(U^T L_Y U\right) 
\end{gathered}
\end{equation}
where a graph Laplacian matrix $L_Y\in {\mathbb{R}^{\text{n}\times n}}$ for $Y$ is indicated by the notation $L_{Y}=G_{Y}-S_{Y}$. $S_{Y}$ is the affinity graph for $Y$, and $G_{Y}$ is the diagonal matrix with $G_{Y}(i,i)=\sum_{j=1}^{n} S_{Y}(i,j)$. A heat kernel has been employed to generate the affinity graph $S_{Y}$. The similarity value of two labels, $\bm{y}_{.i}$ and $\bm{y}_{.j}$, is represented by the element $S_{Y}(i,j)$. $S_{Y}(i,j)$ is defined as follows:
\begin{equation}
S_{Y}(i,j)=\left\{\begin{array}{lc}
\exp \left(-\frac{\left\|\bm{y}_{.i}-\bm{y}_{.j}\right\|^{2}}{\sigma^{2}}\right) &\bm{y}_{.i} \in \mathcal{N}_{q}\left(\bm{y}_{.j}\right) \text { or } \\ & \bm{y}_{.j} \in \mathcal{N}_{q}\left(\bm{y}_{.i}\right) \\
0 & \text { otherwise }
\end{array}\right.
\end{equation}
where the symbol $\sigma$ represents the graph construction parameter, and $\mathcal{N}_{p}\left(\bm{y}_{.j}\right)$ denotes the set of the top $q$ closest neighbours of the label $\bm{y}_{.j}$. Referring to the study\cite{jian2016multi}, the value of $\sigma$ is set to 1.

\subsection{The final objective function of ASLSL}
By combining Eq.~\eqref{equ_framework_wlls} and Eq.~\eqref{equ_framework_C} and introducing $l_{2,1}$-norm sparsity on the projection matrix $Q^{(v)}$, we can obtain the final cost function of ASLSL:
\begin{equation}
\begin{aligned}
&\min_{Q^{(v)}, \alpha^{(v)}, M, U} \sum_{v = 1}^{m} \left(\alpha^{(v)}\right)^{\gamma} \left[ \left\|\left(X^{(v)} - Q^{(v)} U^T\right) S^{(v)}\right\|_F^2 \right. \\
& \left. + \lambda \left\|Y - M U^T\right\|_F^2 + \eta \operatorname{Tr}\left(U^T L_Y U\right) \right. \left.+ \delta \left\|Q^{(v)}\right\|_{2,1} \right] \\
&\text{s.t.} M \geq 0, \quad Q^{(v)} \geq 0, \quad U \geq 0, \\
& 0 \leq \alpha^{(v)} \leq 1, \quad \sum_{v = 1}^{m} \alpha^{(v)} = 1, \quad v = 1, \ldots, m
\end{aligned}
\label{eq_framework}
\end{equation}
where $\lambda$, $\eta$, $\gamma$, and $\delta$ are regularization parameters. The flowchart of ASLSL is shown in Fig.~\ref{Framework_ASLSL}.

\section{Optimization Strategy} \label{Optimization Strategy}
The alternatively iterative update algorithm is adopted to derive solutions for the four variables ($Q^{(v)}$, $U$, $M$, and $\alpha^{(v)}$) in Eq.~\eqref{eq_framework}. The ASLSL algorithm is presented as follows: 

\subsection{Update $Q^{(v)}$ by fixing other variables}
When $U$, $M$, and $\alpha^{(v)}$ are fixed, we remove the irrelevant terms and introduce a Lagrange multiplier $\mathbf{\Psi}$ for $Q^{(v)} \geq 0$. Then, we obtain the following function about $Q^{(v)}$:
\begin{equation}
\begin{aligned}
\min_{Q^{(v)}} \sum_{v=1}^{m} & \left(a^{(v)}\right)^{\gamma} \left[\left\|\left(X^{(v)} - Q^{(v)} U^{\top}\right) S^{(v)}\right\|_{F}^{2} \right. \\
& \left. + \delta \left\|Q^{(v)}\right\|_{2,1}\right] + \operatorname{Tr}\left(\mathbf{\Psi} Q^{(v)}\right)
\end{aligned}
\end{equation}

By taking the partial derivative w.r.t.$Q^{(v)}$, we could get
\begin{equation}
\begin{aligned}
\frac{\partial L\left(Q^{(v)}\right)}{\partial Q^{(v)}} &= \left(a^{(v)}\right)^{\gamma} \left[2 \left(Q^{(v)} U^{\top} S^{(v)} S^{(v)^{T}} U \right. \right. \\
& \left. \left. - X^{(v)} S^{(v)} S^{(v)^{T}} U \right) + 2 \gamma D^{(v)} Q^{(v)}\right] + \mathbf{\Psi}
\end{aligned}
\end{equation}
where $D^{(\mathrm{v})}$ is a diagonal matrix and the element of $D^{(\mathrm{v})}$ is calculated by $D_{i i}^{(\mathrm{v})}=\frac{1}{2 \sqrt{{Q^{(\mathrm{v})}_i}^T Q^{(\mathrm{v})}_i+\epsilon}}(\epsilon \rightarrow 0)$.

According to Karush-Kuhn-Tucker (KKT) complementary condition, $Q^{(\mathrm{v})}$ can be updated as follows:
\begin{equation}
\label{sol_q}
Q^{(v)} \leftarrow Q^{(v)} \odot \frac{X^{(v)} S^{(v)} S^{(v)^{\top}} U}{Q^{(v)} U^{\top} S^{(v)} S^{(v)} U+\gamma D^{(v)} Q^{(v)}}
\end{equation}

\subsection{Update $U$ by fixing other variables}
When $Q^{(v)}$, $M$, and $\alpha^{(v)}$ are fixed, by introducing a Lagrange multiplier $\mathbf{\Theta}$ for $U \geq 0$, we have:
\begin{equation}
\begin{aligned}
&\min_{U} \sum_{v=1}^m \left(\alpha^{(v)}\right)^\gamma \left[ \left\|\left(X^{(v)} - Q^{(v)} U^T\right) S^{(v)}\right\|_F^2 \right. \\
& \left. + \lambda \left\|Y - M U^T\right\|_F^2 \right.\left. + \eta \operatorname{Tr}\left(U^T L_Y U\right) \right] + \operatorname{Tr}\left(\mathbf{\Theta}^T U\right)
\end{aligned}
\end{equation}

Then, the partial derivative w.r.t.$U$ is calculated as:
\begin{equation}
\begin{aligned}
\frac{\partial L\left(Q^{(v)}\right)}{\partial Q^{(v)}} &= \sum_{v=1}^{m} \left(a^{(v)}\right)^{\gamma} [ 2 ( S^{(v)} S^{(v)^{T}} U Q^{(v)^{T}} Q^{(v)} \\
& - S^{(v)} S^{(v)^{T}} X^{(v)^{T}} Q^{(v)} )  + 2 \eta L_{Y} U \\
& + 2 \lambda U M^{T} M - 2 \lambda Y^{T} M ] + \mathbf{\Theta}
\end{aligned}
\end{equation}

Via KKT condition $\mathbf{\Theta}_{ij}U_{ij} = 0$, $U$ can be updated as:
\begin{equation}
\label{sol_u}
U \leftarrow U \odot \frac{\sum_{v=1}^m X^{(v)^T} O \cdot Q^{(v)}+\lambda Y^T M}{\sum_{v=1}^m O \cdot U Q^{(v)^T} Q^{(v)}+ \eta L_Y U+\lambda U M^T M}
\end{equation}
where $O=\left(a^{(v)}\right)^\gamma S^{(v)} S^{(v)^T}$.

\subsection{Update $M$ by fixing other variables}
When $Q^{(v)}$, $U$, and $\alpha^{(v)}$ are fixed, we have the following Lagrange function by removing irrelevant terms and introducing a Lagrange multiplier $\mathbf{\Phi}$ for $M \geq 0$:
\begin{equation}
\min _{M} \sum_{v=1}^{m}\left(a^{(v)}\right)^{\gamma}\left[\lambda\left\|Y-M U^{T}\right\|_{F}^{2}\right]+\operatorname{Tr}(\mathbf{\Phi} M)
\end{equation}

Then, the partial derivative w.r.t.$M$ is calculated as:
\begin{equation}
\frac{\partial L(M)}{\partial M}=\sum_{v=1}^{m}\left(a^{(v)}\right)^{\gamma} \left[\lambda\left(2 M U^{T} U-2 Y U\right)\right]+\mathbf{\Phi} 
\end{equation}

Via KKT condition $\mathbf{\Phi}_{ij}M_{ij} = 0$, $M$ can be updated as:
\begin{equation}
\label{sol_m}
M \leftarrow M \odot \frac{\sum_{v=1}^{m}\left(a^{(v)}\right) \gamma Y U}{\sum_{v=1}^{m}\left(a^{(v)}\right)^{r} M U^{T} U}
\end{equation}

\subsection{Update $\alpha^{(v)}$ by fixing other variables}
When other variables are fixed, we have
\begin{equation}
\begin{aligned}
d^{(v)} = & \left[ \left\| \left(X^{(v)} - Q^{(v)} U^T\right) S^{(v)} \right\|_F^2 + \lambda \left\| Y - M U^T \right\|_F^2 \right. \\
& \left. + \eta \operatorname{Tr}\left(U^T L_Y U\right) + \delta \left\| Q^{(v)} \right\|_{2,1} \right]
\end{aligned}
\end{equation}

Then, the optimization problem of $\alpha^{(v)}$ is changed to
\begin{equation}
\min _{a^{(v)}} \sum_{v=1}^{m}\left(a^{(v)}\right)^{\gamma} d^{(v)} \quad \text { s.t. } \sum_{v=1}^{m} \alpha^{(v)}=1,0 \leq \alpha^{(v)} \leq 1 
\end{equation}

By introducing a Lagrange multiplier $\varphi$ for $\sum_{v=1}^{m} \alpha^{(v)}=1$, we have the following Lagrange function:
\begin{equation}
L\left(a^{(v)}\right)=\sum_{v=1}^{m}\left(a^{(v)}\right)^{\gamma} d^{(v)}-\varphi\left[\sum_{v=1}^{m} a^{(v)}-1\right] 
\end{equation}

The partial derivative w.r.t.$\alpha^{(v)}$ is calculated as:
\begin{equation}
\frac{\partial L\left(\alpha^{(v)}\right)}{\partial \alpha^{(v)}}=\gamma\left(\alpha^{(v)}\right)^{\gamma-1} d^{(v)}-\varphi
\end{equation}

Set $\frac{\partial \mathcal{L}\left(\alpha^{(v)}\right)}{\partial \alpha^{(v)}}= 0$, we have
\begin{equation}
\label{sol_alpha}
\alpha^{(v)}=\frac{\left(d^{(v)}\right)^{\frac{1}{1-\gamma}}}{\sum_{v=1}^{m}\left(d^{(v)}\right)^{\frac{1}{1-\gamma}}}
\end{equation}
\begin{algorithm}[h]
\caption{Adaptive Shared Latent Structure Learning}
\label{ASLSL}
\begin{algorithmic}[1]
\Require Multi-modal physiological signal feature data $X$, multi-dimensional emotional label matrix $Y$, and modal-missing indicator $S$.
\Ensure Return ranked physiological features.
\State Initial $Q^{(v)}$ , $U$, $M$, and $\alpha^{(v)}$ randomly.
\Repeat
\State Update $Q^{(v)}$ via Eq.~\eqref{sol_q};
\State Update $U$ via Eq.~\eqref{sol_u};
\State Update $M$ via Eq.~\eqref{sol_m};
\State Update $\alpha^{(v)}$ via Eq.~\eqref{sol_alpha};
\Until{Convergence;}
\State \Return $Q^{(v)}$ for physiological feature selection.
\State Sort the physiological features by $ \|\bm{q}^{(v)}_{i}\|_{2}$;
\end{algorithmic}
\end{algorithm}

Algorithm~\ref{ASLSL} provides the specific optimization steps for Eq.~\eqref{eq_framework}. The importance of each physiological feature in the multi-dimensional emotion recognition can be evaluated by $Q^{(v)}$. Ultimately, the multi-modal physiological feature subset with informative features is obtained.

\section{Experiments} 
\subsection{Dataset description}
A comprehensive evaluation was conducted using two multi-dimensional emotion datasets, DREAMER \cite{DREAMER2018jbhi} and DEAP \cite{koelstra2011deap}, which encompass multi-modal physiological signal data, to assess the performance of ASLSL. Both datasets adopted the valence-arousal-dominance framework to represent the subject's emotional state. Detailed descriptions of the experimental setup can be found in \cite{DREAMER2018jbhi} and \cite{koelstra2011deap}. The experimental process involved applying a band-pass filter with a frequency range from 1 to 50 Hz to remove noise from the EEG recordings. Following this, independent component analysis was employed to mitigate artifacts from the multi-modal physiological signal data.

\subsection{Feature extraction}
Based on previous research on physiological signal features for emotion recognition \cite{jenke2014taffc,xu2020fsorer,DREAMER2018jbhi,koelstra2011deap}, thirteen distinct EEG features were extracted. These features encompass the absolute power, the beta-to-theta absolute power ratio, C0 complexity, higher-order crossing, non-stationary index, Shannon entropy, spectral entropy, differential entropy, differential asymmetry, rational asymmetry, the instantaneous phase of the Hilbert-transformed intrinsic mode functions, and the amplitude of the Hilbert-transformed intrinsic mode functions. 

For ECG data, we extracted thirteen distinct features related to heart rate and variability. These features encompass heart rate, mean, median, standard deviation, minimum, maximum, range of each segment of the PQRST complexes, as well as the difference between consecutive RR intervals, power spectral density for low frequency (LF) and high frequency (HF), the LF to HF ratio, SHE, and total power \cite{DREAMER2018jbhi}. 

Finally, four distinct EOG features were extracted: eye blinking rate, signal energy, mean of the signal, and signal variance \cite{koelstra2011deap}. A comprehensive description of these physiological features is available in \cite{Duan2013DE,jenke2014taffc,GRMOR2021taffc,DREAMER2018jbhi,koelstra2011deap}. The total dimensionality of these multi-modal physiological signal features are 676 for DREAMER and 1764 for DEAP.

\subsection{Experimental setup}
\textbf{Comparative methods:} To thoroughly assess the performance of ASLSL in multi-dimensional affective computing, seventeen advanced feature selection methods were compared. The comparative methods include:

(1) Four feature selection methods commonly used in the BCI applications: mRMR \cite{peng2005mRMR}, RFS \cite{nie2010RFS}, FSOR \cite{xu2020fsorer}, and GRMOR \cite{GRMOR2021taffc}.

(2) Nine popular multi-label feature selection approaches: RPMFS \cite{cai2013exact},  SCLS \cite{lee2017scls},  MDFS\cite{zhang2019manifold},  MGFS \cite{hashemi2020mgfs}, MFS-MCDM \cite{hashemi2020mfs}, GRRO \cite{zhang2020multilabel}, SDFS \cite{wang2020discriminative}, SLMDS \cite{li2023robust}, and RFSFS \cite{li2023multi}. 

(3) Four advanced multi-view multi-label feature selection methods: MSFS \cite{zhang2020multi}, DHLI \cite{hao2024double}, UGRFS \cite{hao2025uncertainty}, and EF$^2$FS \cite{hao2025embedded}. 

\textbf{Experimental details:} The multi-modal physiological data were categorized into low and high values based on self-assessed scores for each affective dimension, with a threshold value of five. Multi-label k-nearest neighbor (ML-KNN) \cite{ZHANG2007MLKNN} was employed as the base classifier, with the number of nearest neighbors set to 10 and the smoothing parameter set to 1. Seventy percent of the participants were randomly selected for the training set, while the remaining thirty percent constituted the test set. A cross-subject experimental design was utilized to ensure unbiased results. To minimize bias, 50 separate and independent experiments were conducted, and the average results were used as the final measure. It is noteworthy that physiological feature extraction was performed on the entire trial as a single sample, without subdividing trials into smaller segments to augment the sample size for the experiments.

The strategy described in \cite{liu2018late} was adopted to simulate an incomplete multi-modal physiological data scenario by removing specific percentages of instances from each modality to represent partial data absence. The missing data ratio ranged from 10\% to 50\%, in increments of 10\%. Approximately 10\% of all features were selected using the feature selection methods. The parameters ($\lambda$, $\eta$, and $\delta$) were adjusted within the range of $10^{-3}$ to $10^3$ with a step size of $10^1$. The parameter $\gamma$ varied within the set $\left\{2,3,4,5,6,7,8,9\right\}$. The parameters of comparative methods were set as described in the corresponding references.

\textbf{Performance metrics:} To evaluate the effectiveness of multi-dimensional affect computing, six metrics were utilized. These metrics included two label-based measures (micro-F1 (MI) and macro-F1 (MA)) and four example-based metrics: Hamming loss (HL), ranking loss (RL), average precision (AP), and coverage (CV). For CV, HL, and RL, lower values indicate superior performance, ideally approaching zero. For MA, MI, and AP, higher values correspond to better classification outcomes, theoretically nearing the maximum value of 1. These metrics collectively quantify model performance across dimensions such as label correlation, classification accuracy, and generalization capability, forming the core evaluation framework for multi-label tasks. For further details, refer to \cite{zhang2019manifold}.



\begin{figure}[!t]
\centering
\includegraphics[width=0.5\textwidth]{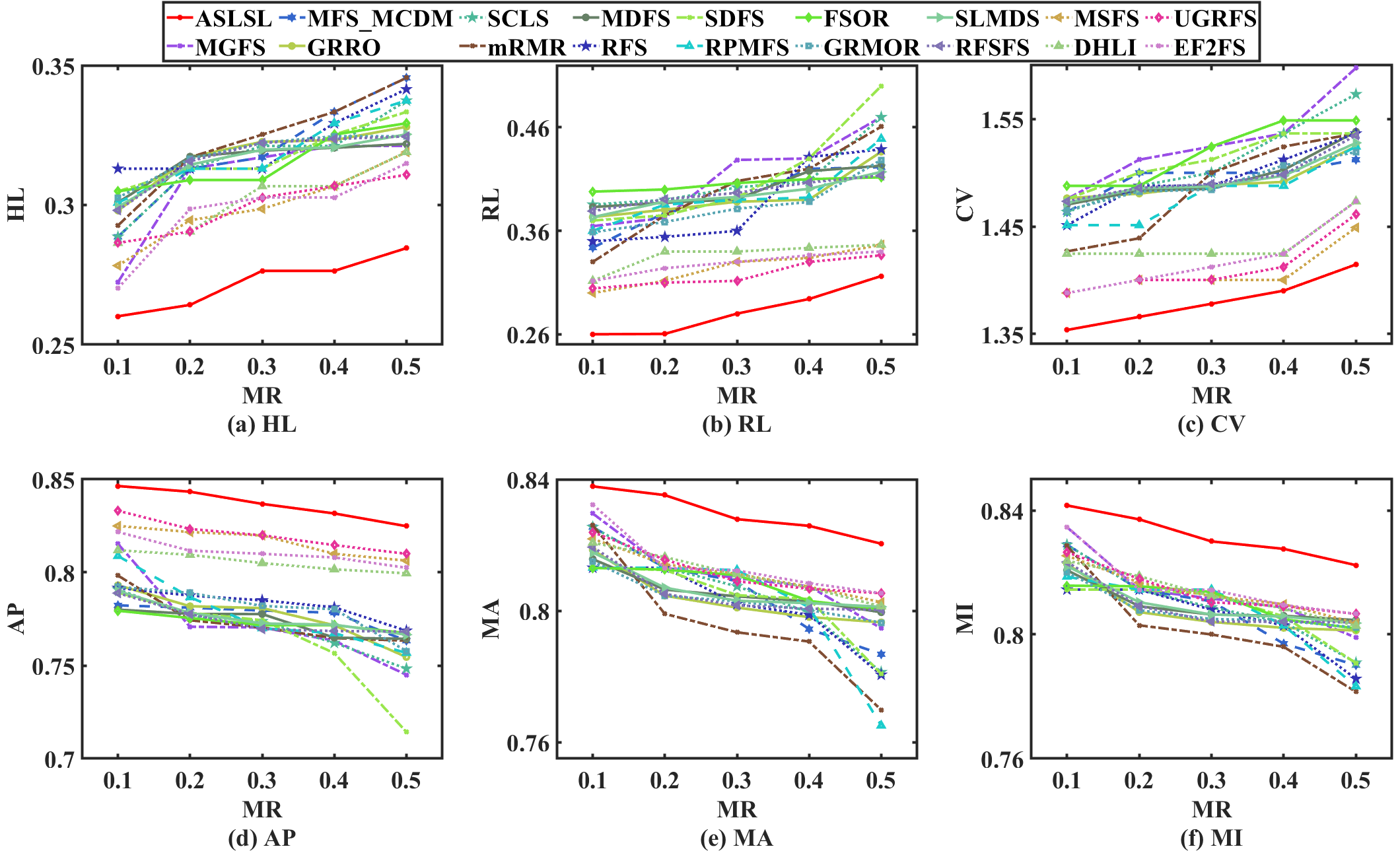}
\caption{Multi-dimensional emotion recognition performance of various missing ratios (MR) on DREAMER.}\label{Results_index_dreamer}
\end{figure}

\begin{figure}[!t]
\centering
\includegraphics[width=0.5\textwidth]{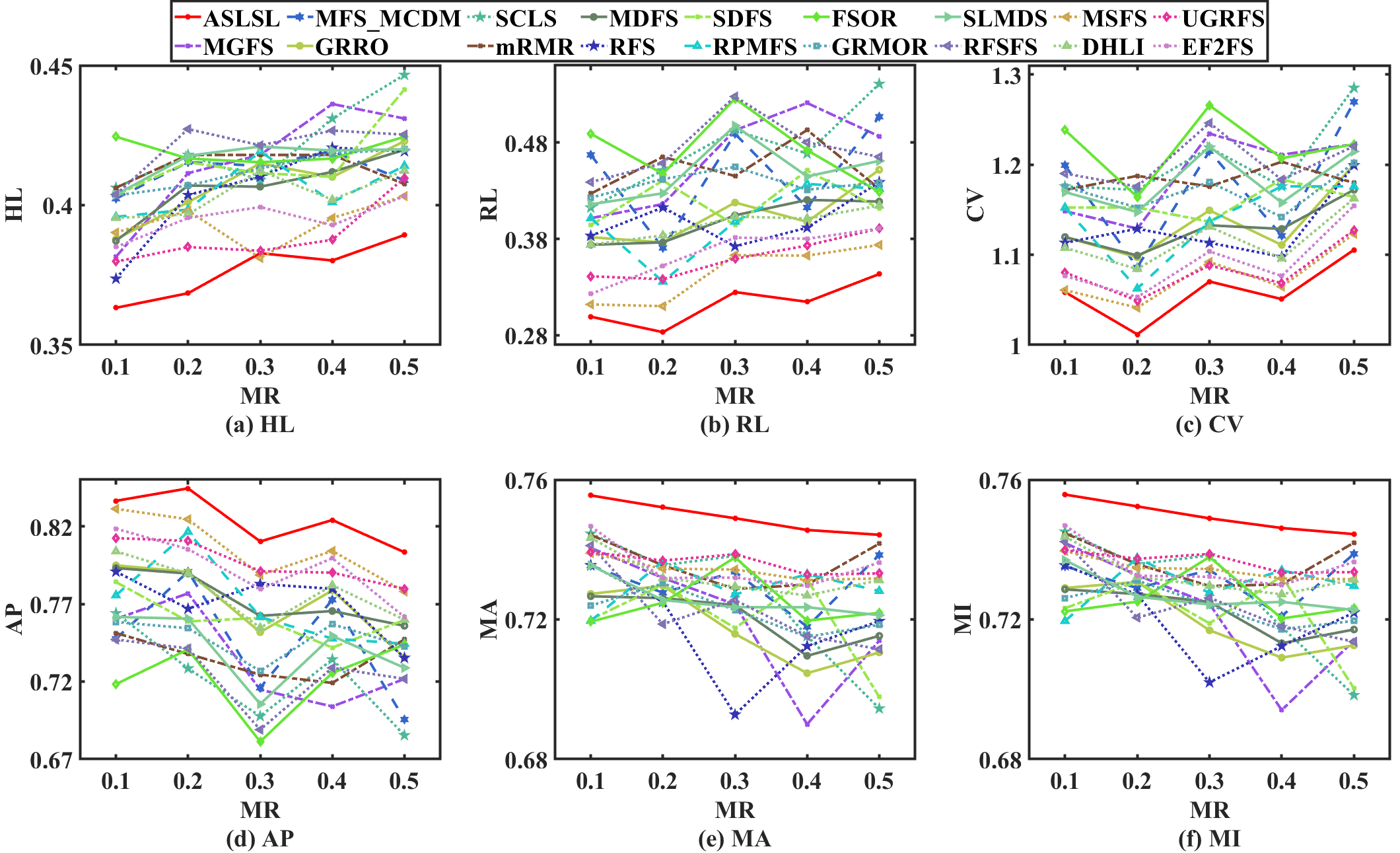}
\caption{Multi-dimensional emotion recognition performance of various missing ratios (MR) on DEAP.}\label{Results_index_deap}
\end{figure}

\subsection{Emotion recognition performance with incomplete physiological data}
Fig.~\ref{Results_index_deap} and Fig.~\ref{Results_index_dreamer} illustrate the comparative performance results for the DEAP and DREAMER dataset, respectively. The horizontal axis in each figure represents the missing ratio of multi-modal physiological signals, while the vertical axis shows the performance of each method according to the given indicator. In all the subgraphs, ASLSL method is denoted by the red line. As depicted in Fig.~\ref{Results_index_deap} and Fig.~\ref{Results_index_dreamer}, ASLSL consistently outperforms the other seventeen methods across all indicators and various levels of missing data.

To further evaluate the performance of the eighteen feature selection methods, a Friedman test was conducted. The significance level ($\alpha$) was set to 0.05. The results of this statistical significance test are summarized in Table~\ref{tab:friedman}, which indicates that the null hypothesis is rejected, signifying significant differences in performance among the eighteen feature selection methods for multi-dimensional emotion recognition with incomplete multi-modal physiological signal data. 

\begin{table}[!t]\small
\setlength{\abovecaptionskip}{0.cm}
\setlength{\belowcaptionskip}{-0.cm}
\begin{center}
\small
{
\begin{tabular}{lcc}
\hline\hline
Evaluation metric     & $F_{F}$             & Critical value  \\ \hline
Ranking loss          & 13.282              & \multicolumn{1}{c}{\multirow{7}*{$\approx$ 2.272}}             \\
Coverage              & 16.986             &            \\
Hamming loss          & 10.098             &            \\
Average precision     & 14.954             &           \\
Macro-F1              & 4.673            &            \\
Micro-F1              & 4.694             &            \\
\hline\hline
\end{tabular}}
\end{center}
\caption{The Friedman test results (significance level $\alpha = 0.05$).}\label{tab:friedman}
\vspace{-0.14cm}
\end{table}

\begin{table}[!t]\small
\begin{center}
\small
{
\begin{tabular}{ccc}
\hline\hline
\multicolumn{1}{l}{\multirow{2}*{Conditions}}         & \multicolumn{2}{c}{\multirow{1}*{Average precision}} \\  \cline{2-3}
                          & DEAP           & DREAMER                \\\hline
w/o ASLSL                  & 0.72           & 0.76              \\
w/o GMR                   & 0.76           & 0.79              \\
w/o AMWE                  & 0.75           & 0.77              \\
Our method                & 0.82           & 0.84               \\ \hline\hline
\end{tabular}}
\end{center}
\caption{The results of ablation experiments (w/o, ASLSL, GMR, and AMWE denote without, adaptive shared latent structure learning, graph-based manifold regularizer, and adaptive modal weight evaluation, respectively).}\label{tab:acml}
\end{table}

\begin{figure}[!t]
\centering
\subfigure[DEAP]{\label{Conv_deap}\includegraphics[width=0.23\textwidth]{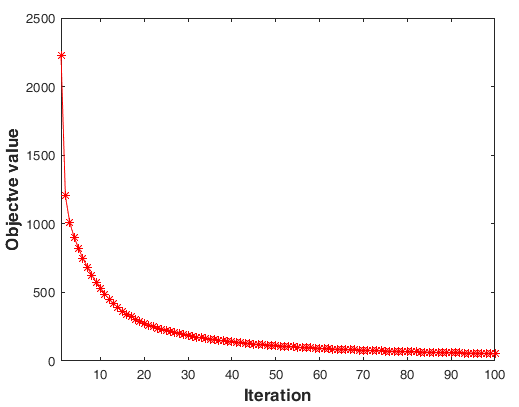}}
\hspace{0.0cm}
\subfigure[DREAMER]{\label{Conv_dreamer}\includegraphics[width=0.23\textwidth]{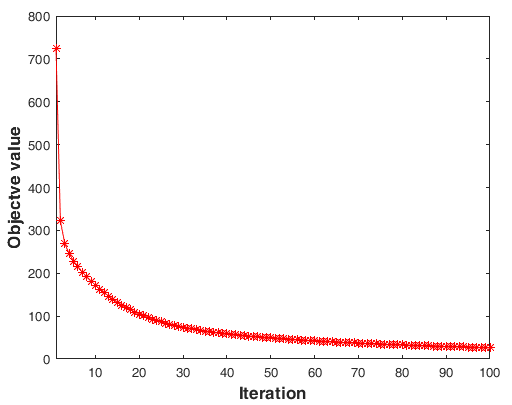}}
\hspace{0.0cm}
\caption{Convergence verification of ASLSL.}
\vspace{-0.14cm}
\label{conv_res}
\end{figure}
\subsection{Ablation experiments}
To evaluate the contributions of each component within the proposed EEG feature selection model, we conducted ablation experiments. The ASLSL framework comprises three essential modules, and we systematically removed each module to assess its impact. Table~\ref{tab:acml} reveals that adaptive shared latent structure learning is pivotal for addressing the challenges posed by incomplete data, as it facilitates the exploration of a common space for multi-modal physiological signals and multi-dimensional emotions. The other modules are instrumental in preserving local geometric structures and evaluating the significance of each model.

\subsection{Convergence analysis}
Additionally, we performed an analysis to assess the convergence speed of the iterative optimization algorithm proposed. Fig.~\ref{conv_res} illustrates the convergence trajectories of the objective function for both the DREAMER and DEAP datasets. The tradeoff parameters ($\lambda$, $\eta$, and $\delta$) were all set to 1 and $r$ was set to 2. As shown in Fig.~\ref{conv_res}, the ASLSL algorithm exhibits rapid convergence, highlighting the effectiveness of our optimization algorithm.

\begin{figure}[!t]
\centering
\subfigure[$\gamma$]{\label{bar3_lambda}\includegraphics[width=0.2\textwidth]{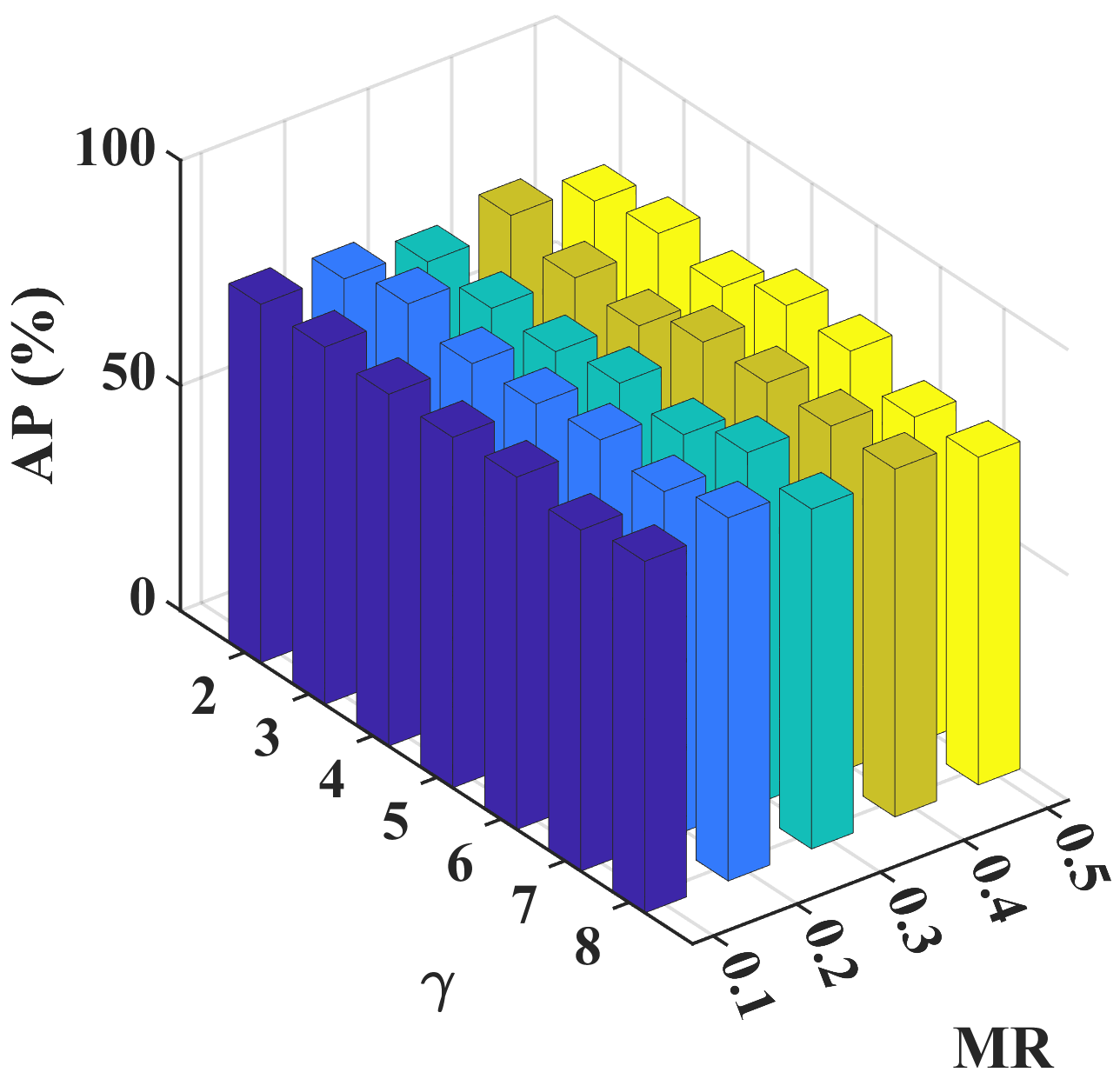}}
\hspace{0.0cm}
\subfigure[$\eta$]{\label{bar3_eta}\includegraphics[width=0.2\textwidth]{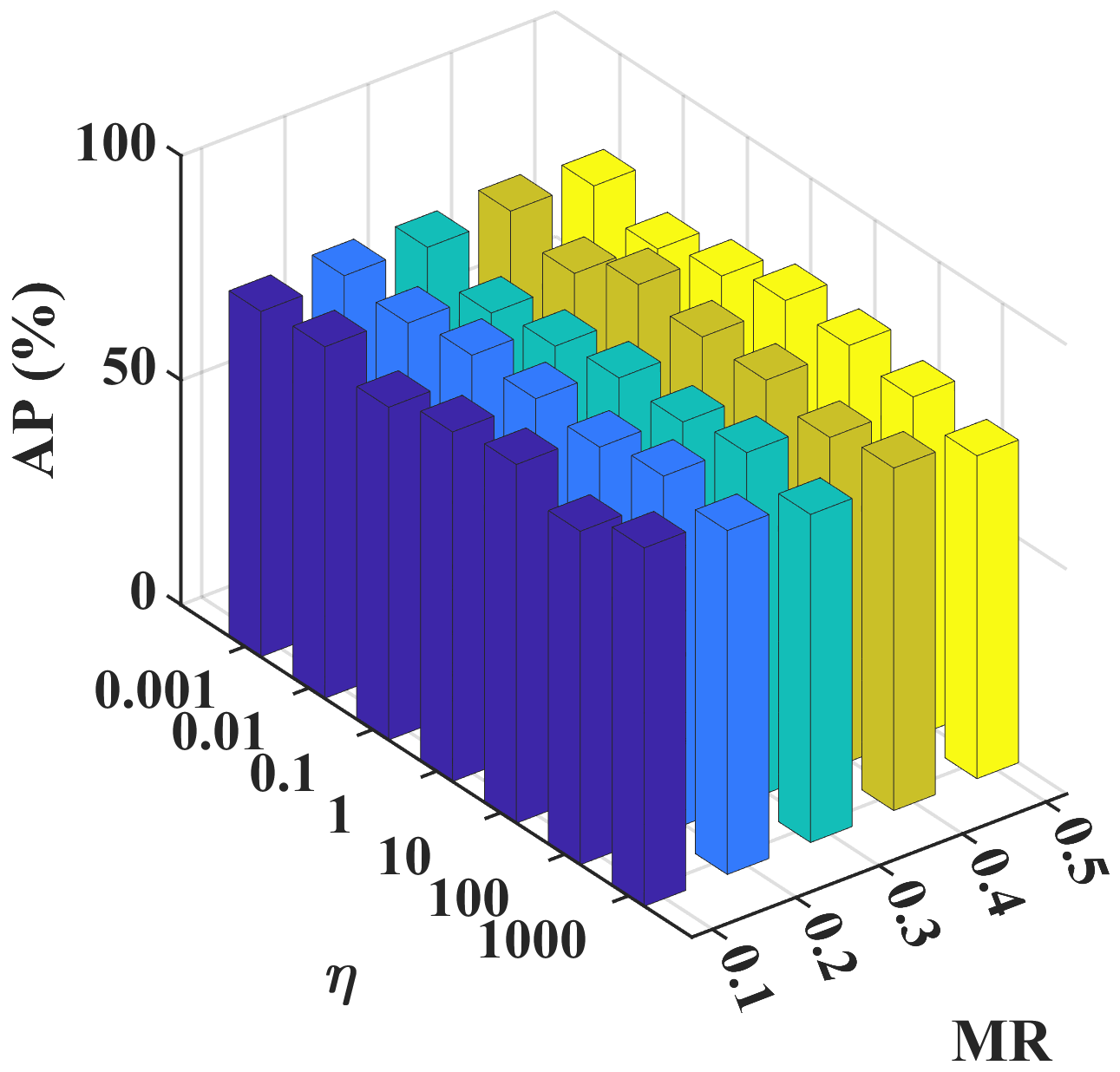}}
\hspace{0.0cm}
\subfigure[$\lambda$]{\label{bar3_mu}\includegraphics[width=0.2\textwidth]{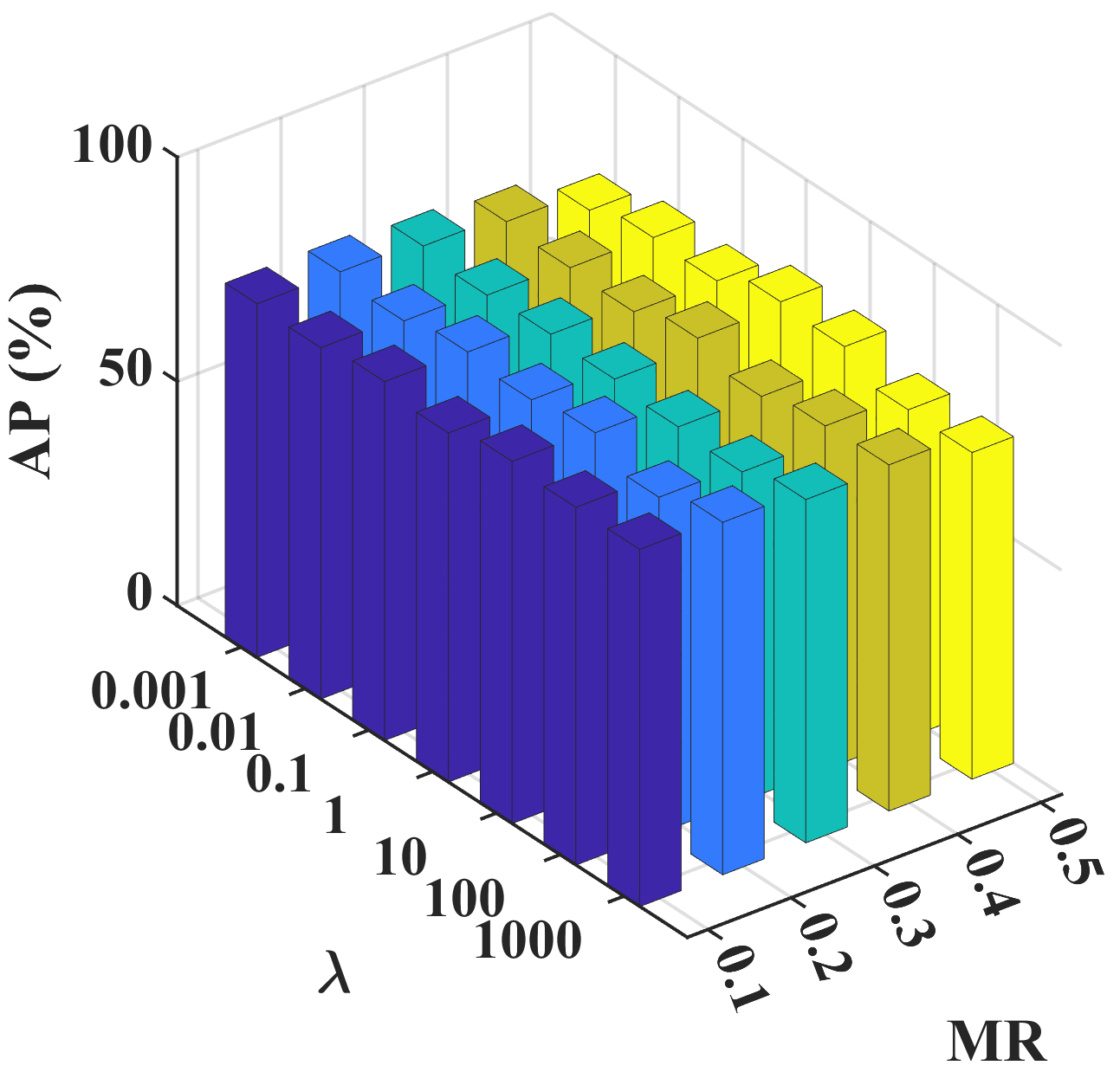}}
\hspace{0.0cm}
\subfigure[$\delta$]{\label{bar3_delta}\includegraphics[width=0.2\textwidth]{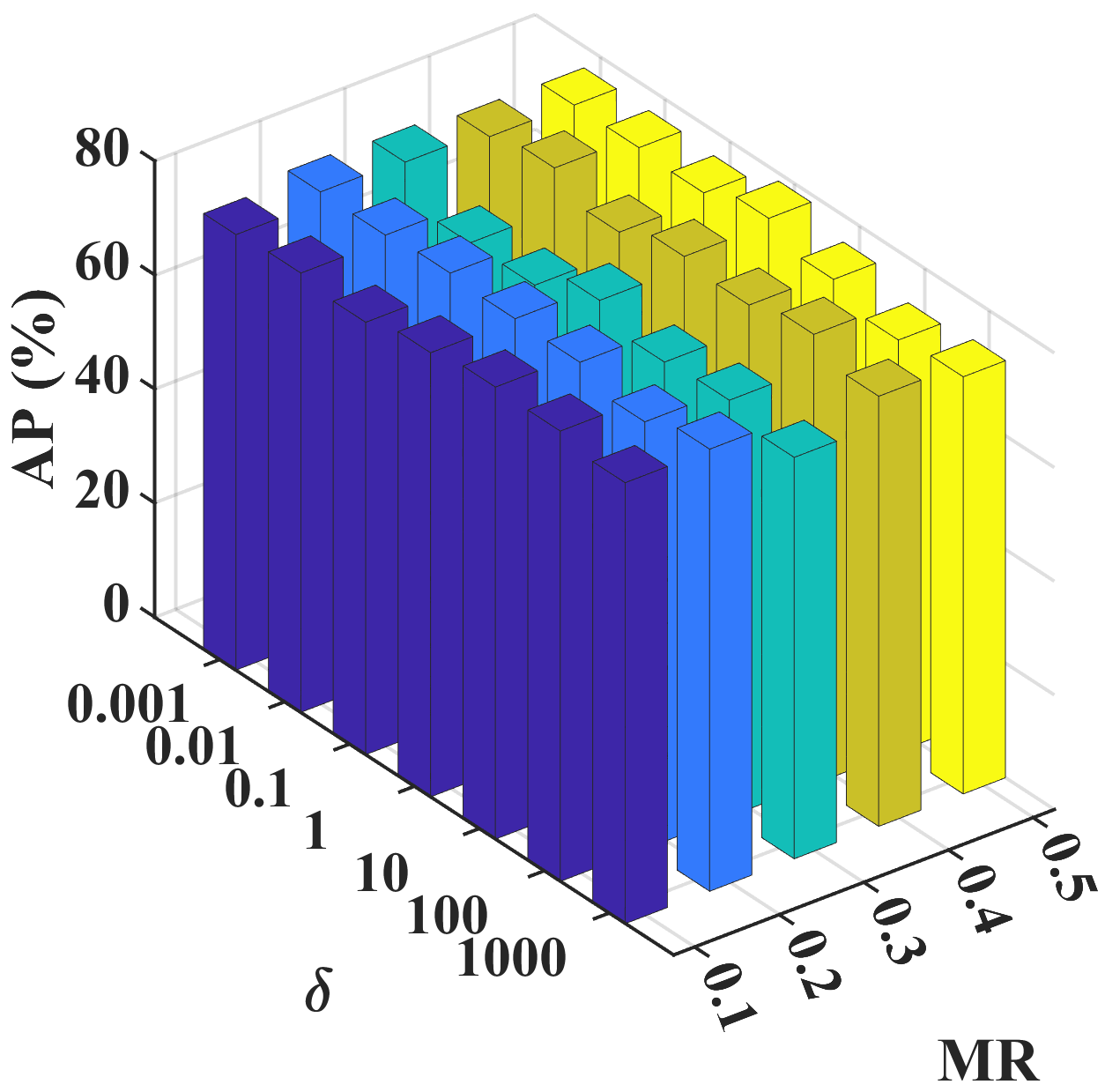}}
\hspace{0.0cm}
\caption{The parameter sensitivity results on DEAP.}
\vspace{-0.14cm}
\label{Results_bar3_deap}
\end{figure}

\begin{table}[!t]\small
\begin{center}
\small
{
\begin{tabular}{ccc}
\hline\hline
Methods         &  DEAP       & DREAMER  \\\hline
MGFS	&0.23	&0.35     \\
MFS-MCDM	&0.21	&0.19 \\
SCLS	&15.93	&1.27\\
mRMR	&0.63	&0.09\\
RFS	&8.22	&0.59\\
SDFS	&1.01	&0.16\\
RPMFS	&4.94	&0.98\\
FSOR	&359.77	&54.07\\
GRMOR	&35.91	&7.22\\
GRRO	&15.62	&1.47\\
MDFS	&23.80	&2.07\\
SLMDS	&0.71	&0.08\\
RFSFS	&1.62	&0.41\\
MSFS    &1.67	&1.13\\
DHLI    &0.60	&0.17 \\   
UGRFS   &104.23	&4.78 \\ 
EF2FS   &1.42	&0.34 \\
\textbf{ASLSL}            & \textbf{7.13}          & \textbf{1.07}        \\ \hline\hline
\end{tabular}}
\end{center}
\hspace{0.0cm}
\caption{The computational time results (seconds).}\label{tab:avetime}
\vspace{-0.14cm}
\end{table}

\subsection{Parameter sensitivity}
This section presents a sensitivity analysis of ASLSL concerning these four parameters. For each analysis, one parameter was held constant at 0.1 ($\gamma = 2$), while the others were varied within the specified range. Due to space limitations, we provide only the sensitivity analysis results for the DEAP dataset. The 3-D histograms, depicting the AP values, are presented in Fig.\ref{Results_bar3_deap}. As shown in Fig.\ref{Results_bar3_deap}, the AP values remain relatively stable across variations in two parameters, indicating that the performance of ASLSL is not sensitive to these balance parameters.

\subsection{Computational cost}
The computational time required by each method was assessed. The implementation was carried out using MATLAB (MathWorks Inc., Novi, MI, USA) and was executed on a computer running Microsoft Windows $11 \times 64$, featuring an Intel i5-13500H 2.60 GHz CPU and 16.00 GB of RAM. The results for average computational time are summarized in Table~\ref{tab:avetime}. As illustrated in Table~\ref{tab:avetime}, ASLSL has superior performance in recognizing emotions from incomplete data while maintaining a relatively low computational cost.

\section{Conclusions}\label{Conclusion}
We introduce a novel approach for multi-modal physiological signal feature selection specifically designed to handle incomplete physiological signal data in emotion recognition tasks. This method utilizes adaptive shared latent structure learning to investigate a unified space for multi-modal physiological signals and multi-dimensional emotions, effectively mitigating the effects of missing information and enabling the selection of informative features. Furthermore, we propose an efficient iterative algorithm to address the ASLSL optimization problem. Experimental results demonstrate that ASLSL markedly surpasses seventeen advanced methods in performance.

\bibliography{aaai2026}

\begin{thebibliography}{39}
\providecommand{\natexlab}[1]{#1}

\bibitem[{Becker et~al.(2017)Becker, Fleureau, Guillotel, Wendling, Merlet, and
  Albera}]{Becker2020HRE}
Becker, H.; Fleureau, J.; Guillotel, P.; Wendling, F.; Merlet, I.; and Albera,
  L. 2017.
\newblock Emotion recognition based on high-resolution EEG recordings and
  reconstructed brain sources.
\newblock \emph{IEEE Transactions on Affective Computing}, 11(2): 244--257.

\bibitem[{Cai, Nie, and Huang(2013)}]{cai2013exact}
Cai, X.; Nie, F.; and Huang, H. 2013.
\newblock Exact top-k feature selection via $l_{2,1}$-norm constraint.
\newblock In \emph{Twenty-third international joint conference on artificial
  intelligence}.

\bibitem[{Duan, Zhu, and Lu(2013)}]{Duan2013DE}
Duan, R.; Zhu, J.; and Lu, B. 2013.
\newblock Differential entropy feature for EEG-based emotion classification.
\newblock In \emph{2013 6th international IEEE/EMBS conference on neural
  engineering (NER)}, 81--84. IEEE.

\bibitem[{Ezzameli and Mahersia(2023)}]{ezzameli2023emotion}
Ezzameli, K.; and Mahersia, H. 2023.
\newblock Emotion recognition from unimodal to multimodal analysis: A review.
\newblock \emph{Information Fusion}, 101847.

\bibitem[{Geetha et~al.(2024)Geetha, Mala, Priyanka, and
  Uma}]{geetha2024multimodal}
Geetha, A.; Mala, T.; Priyanka, D.; and Uma, E. 2024.
\newblock Multimodal Emotion Recognition with deep learning: advancements,
  challenges, and future directions.
\newblock \emph{Information Fusion}, 105: 102218.

\bibitem[{Hao, Gao, and Hu(2025)}]{hao2025embedded}
Hao, P.; Gao, W.; and Hu, L. 2025.
\newblock Embedded feature fusion for multi-view multi-label feature selection.
\newblock \emph{Pattern Recognition}, 157: 110888.

\bibitem[{Hao, Liu, and Gao(2024)}]{hao2024double}
Hao, P.; Liu, K.; and Gao, W. 2024.
\newblock Double-layer hybrid-label identification feature selection for
  multi-view multi-label learning.
\newblock In \emph{Proceedings of the AAAI Conference on Artificial
  Intelligence}, volume~38, 12295--12303.

\bibitem[{Hao, Liu, and Gao(2025)}]{hao2025uncertainty}
Hao, P.; Liu, K.; and Gao, W. 2025.
\newblock Uncertainty-Aware Global-View Reconstruction for Multi-View
  Multi-Label Feature Selection.
\newblock In \emph{Proceedings of the AAAI Conference on Artificial
  Intelligence}, volume~39, 17068--17076.

\bibitem[{Hashemi, Dowlatshahi, and
  Nezamabadi-pour(2020{\natexlab{a}})}]{hashemi2020mfs}
Hashemi, A.; Dowlatshahi, M.~B.; and Nezamabadi-pour, H. 2020{\natexlab{a}}.
\newblock MFS-MCDM: Multi-label feature selection using multi-criteria decision
  making.
\newblock \emph{Knowledge-Based Systems}, 206: 106365.

\bibitem[{Hashemi, Dowlatshahi, and
  Nezamabadi-pour(2020{\natexlab{b}})}]{hashemi2020mgfs}
Hashemi, A.; Dowlatshahi, M.~B.; and Nezamabadi-pour, H. 2020{\natexlab{b}}.
\newblock MGFS: A multi-label graph-based feature selection algorithm via
  PageRank centrality.
\newblock \emph{Expert Systems with Applications}, 142: 113024.

\bibitem[{Hou et~al.(2011)Hou, Nie, Yi, and Wu}]{hou2011JELSR}
Hou, C.; Nie, F.; Yi, D.; and Wu, Y. 2011.
\newblock Feature selection via joint embedding learning and sparse regression.
\newblock In \emph{Twenty-Second international joint conference on Artificial
  Intelligence}, 1324--1329.

\bibitem[{Jenke, Peer, and Buss(2014)}]{jenke2014taffc}
Jenke, R.; Peer, A.; and Buss, M. 2014.
\newblock Feature Extraction and Selection for Emotion Recognition from EEG.
\newblock \emph{IEEE Transactions on Affective Computing}, 5(3): 327--339.

\bibitem[{Jian et~al.(2016)Jian, Li, Shu, and Liu}]{jian2016multi}
Jian, L.; Li, J.; Shu, K.; and Liu, H. 2016.
\newblock Multi-label informed feature selection.
\newblock In \emph{The twenty-fifth International Joint Conference on
  Artificial Intelligence}, 1627--1633.

\bibitem[{Katsigiannis and Ramzan(2018)}]{DREAMER2018jbhi}
Katsigiannis, S.; and Ramzan, N. 2018.
\newblock DREAMER: A Database for Emotion Recognition Through EEG and ECG
  Signals From Wireless Low-cost Off-the-Shelf Devices.
\newblock \emph{IEEE Journal of Biomedical and Health Informatics}, 22(1):
  98--107.

\bibitem[{Khare et~al.(2023)Khare, Blanes-Vidal, Nadimi, and
  Acharya}]{khare2023emotion}
Khare, S.~K.; Blanes-Vidal, V.; Nadimi, E.~S.; and Acharya, U.~R. 2023.
\newblock Emotion recognition and artificial intelligence: A systematic review
  (2014--2023) and research recommendations.
\newblock \emph{Information Fusion}, 102019.

\bibitem[{Koelstra et~al.(2011)Koelstra, Muhl, Soleymani, Lee, Yazdani,
  Ebrahimi, Pun, Nijholt, and Patras}]{koelstra2011deap}
Koelstra, S.; Muhl, C.; Soleymani, M.; Lee, J.-S.; Yazdani, A.; Ebrahimi, T.;
  Pun, T.; Nijholt, A.; and Patras, I. 2011.
\newblock Deap: A database for emotion analysis; using physiological signals.
\newblock \emph{IEEE transactions on affective computing}, 3(1): 18--31.

\bibitem[{Lee and Kim(2017)}]{lee2017scls}
Lee, J.; and Kim, D.-W. 2017.
\newblock SCLS: Multi-label feature selection based on scalable criterion for
  large label set.
\newblock \emph{Pattern Recognition}, 66: 342--352.

\bibitem[{Li et~al.(2017)Li, Cheng, Wang, Morstatter, Trevino, Tang, and
  Liu}]{li2017feature}
Li, J.; Cheng, K.; Wang, S.; Morstatter, F.; Trevino, R.~P.; Tang, J.; and Liu,
  H. 2017.
\newblock Feature selection: A data perspective.
\newblock \emph{ACM computing surveys (CSUR)}, 50(6): 1--45.

\bibitem[{Li et~al.(2022)Li, Zhang, Tiwari, Song, Hu, Yang, Zhao, Kumar, and
  Marttinen}]{li2022eeg}
Li, X.; Zhang, Y.; Tiwari, P.; Song, D.; Hu, B.; Yang, M.; Zhao, Z.; Kumar, N.;
  and Marttinen, P. 2022.
\newblock EEG based emotion recognition: A tutorial and review.
\newblock \emph{ACM Computing Surveys}, 55(4): 1--57.

\bibitem[{Li, Hu, and Gao(2023{\natexlab{a}})}]{li2023multi}
Li, Y.; Hu, L.; and Gao, W. 2023{\natexlab{a}}.
\newblock Multi-label feature selection via robust flexible sparse
  regularization.
\newblock \emph{Pattern Recognition}, 134: 109074.

\bibitem[{Li, Hu, and Gao(2023{\natexlab{b}})}]{li2023robust}
Li, Y.; Hu, L.; and Gao, W. 2023{\natexlab{b}}.
\newblock Robust sparse and low-redundancy multi-label feature selection with
  dynamic local and global structure preservation.
\newblock \emph{Pattern Recognition}, 134: 109120.

\bibitem[{Liu et~al.(2018{\natexlab{a}})Liu, Zhu, Li, Wang, Tang, Yin, Shen,
  Wang, and Gao}]{liu2018late}
Liu, X.; Zhu, X.; Li, M.; Wang, L.; Tang, C.; Yin, J.; Shen, D.; Wang, H.; and
  Gao, W. 2018{\natexlab{a}}.
\newblock Late fusion incomplete multi-view clustering.
\newblock \emph{IEEE transactions on pattern analysis and machine
  intelligence}, 41(10): 2410--2423.

\bibitem[{Liu et~al.(2018{\natexlab{b}})Liu, Xie, Wu, Cao, Li, and
  Li}]{liu2018electroencephalogram}
Liu, Z.-T.; Xie, Q.; Wu, M.; Cao, W.-H.; Li, D.-Y.; and Li, S.-H.
  2018{\natexlab{b}}.
\newblock Electroencephalogram emotion recognition based on empirical mode
  decomposition and optimal feature selection.
\newblock \emph{IEEE Transactions on Cognitive and Developmental Systems},
  11(4): 517--526.

\bibitem[{Nie et~al.(2010)Nie, Huang, Cai, and Ding}]{nie2010RFS}
Nie, F.; Huang, H.; Cai, X.; and Ding, C.~H. 2010.
\newblock Efficient and robust feature selection via joint $l_{2,1}$-norms
  minimization.
\newblock In \emph{Advances in neural information processing systems},
  1813--1821.

\bibitem[{Peng, Long, and Ding(2005)}]{peng2005mRMR}
Peng, H.; Long, F.; and Ding, C. 2005.
\newblock Feature selection based on mutual information: criteria of
  max-dependency, max-relevance, and min-redundancy.
\newblock \emph{IEEE Transactions on Pattern Analysis \& Machine Intelligence},
  (8): 1226--1238.

\bibitem[{Torres-Valencia, {\'A}lvarez-L{\'o}pez, and
  Orozco-Guti{\'e}rrez(2017)}]{torres2017svm}
Torres-Valencia, C.; {\'A}lvarez-L{\'o}pez, M.; and Orozco-Guti{\'e}rrez,
  {\'A}. 2017.
\newblock SVM-based feature selection methods for emotion recognition from
  multimodal data.
\newblock \emph{Journal on Multimodal User Interfaces}, 11: 9--23.

\bibitem[{Wang et~al.(2020{\natexlab{a}})Wang, Wu, Zhang, Xu, Zhang, Wu, and
  Coleman}]{wang2020emotion}
Wang, F.; Wu, S.; Zhang, W.; Xu, Z.; Zhang, Y.; Wu, C.; and Coleman, S.
  2020{\natexlab{a}}.
\newblock Emotion recognition with convolutional neural network and EEG-based
  EFDMs.
\newblock \emph{Neuropsychologia}, 107506.

\bibitem[{Wang, Li, and Cui(2024)}]{wang2024incomplete}
Wang, Y.; Li, Y.; and Cui, Z. 2024.
\newblock Incomplete multimodality-diffused emotion recognition.
\newblock \emph{Advances in Neural Information Processing Systems}, 36.

\bibitem[{Wang, Zhang, and Di(2024)}]{wang2024research}
Wang, Y.; Zhang, B.; and Di, L. 2024.
\newblock Research Progress of EEG-Based Emotion Recognition: A Survey.
\newblock \emph{ACM Computing Surveys}, 56(11): 1--49.

\bibitem[{Wang et~al.(2020{\natexlab{b}})Wang, Nie, Tian, Wang, and
  Li}]{wang2020discriminative}
Wang, Z.; Nie, F.; Tian, L.; Wang, R.; and Li, X. 2020{\natexlab{b}}.
\newblock Discriminative feature selection via a structured sparse subspace
  learning module.
\newblock In \emph{Proc. Twenty-Ninth Int. Joint Conf. Artif. Intell.},
  3009--3015.

\bibitem[{Wu et~al.(2023)Wu, Lu, Hu, and Zeng}]{wu2023affective}
Wu, D.; Lu, B.-L.; Hu, B.; and Zeng, Z. 2023.
\newblock Affective brain--computer interfaces (abcis): A tutorial.
\newblock \emph{Proceedings of the IEEE}, 111(10): 1314--1332.

\bibitem[{Xu et~al.(2023)Xu, Jia, Li, Wei, Ye, and Wu}]{GRMOR2021taffc}
Xu, X.; Jia, T.; Li, Q.; Wei, F.; Ye, L.; and Wu, X. 2023.
\newblock EEG Feature Selection via Global Redundancy Minimization for Emotion
  Recognition.
\newblock \emph{IEEE Transactions on Affective Computing}, 14(1): 421--435.

\bibitem[{Xu et~al.(2020)Xu, Wei, Zhu, Liu, and Wu}]{xu2020fsorer}
Xu, X.; Wei, F.; Zhu, Z.; Liu, J.; and Wu, X. 2020.
\newblock Eeg Feature Selection Using Orthogonal Regression: Application to
  Emotion Recognition.
\newblock In \emph{ICASSP 2020 - 2020 IEEE International Conference on
  Acoustics, Speech and Signal Processing (ICASSP)}, 1239--1243.

\bibitem[{Zhang et~al.(2020{\natexlab{a}})Zhang, Lin, Jiang, Li, Tang, and
  Tan}]{zhang2020multilabel}
Zhang, J.; Lin, Y.; Jiang, M.; Li, S.; Tang, Y.; and Tan, K.~C.
  2020{\natexlab{a}}.
\newblock Multi-label Feature Selection via Global Relevance and Redundancy
  Optimization.
\newblock In \emph{the twenty-ninth International Joint Conference on
  Artificial Intelligence}, 2512--2518.

\bibitem[{Zhang et~al.(2019{\natexlab{a}})Zhang, Luo, Li, Zhou, and
  Li}]{zhang2019manifold}
Zhang, J.; Luo, Z.; Li, C.; Zhou, C.; and Li, S. 2019{\natexlab{a}}.
\newblock Manifold regularized discriminative feature selection for multi-label
  learning.
\newblock \emph{Pattern Recognition}, 95: 136--150.

\bibitem[{Zhang et~al.(2020{\natexlab{b}})Zhang, Yin, Chen, and
  Nichele}]{zhang2020emotion}
Zhang, J.; Yin, Z.; Chen, P.; and Nichele, S. 2020{\natexlab{b}}.
\newblock Emotion recognition using multi-modal data and machine learning
  techniques: A tutorial and review.
\newblock \emph{Information Fusion}, 59: 103--126.

\bibitem[{Zhang and Zhou(2007)}]{ZHANG2007MLKNN}
Zhang, M.; and Zhou, Z. 2007.
\newblock ML-KNN: A lazy learning approach to multi-label learning.
\newblock \emph{Pattern Recognition}, 40(7): 2038 -- 2048.

\bibitem[{Zhang et~al.(2019{\natexlab{b}})Zhang, Nie, Li, and
  Wei}]{zhang2019review}
Zhang, R.; Nie, F.; Li, X.; and Wei, X. 2019{\natexlab{b}}.
\newblock Feature selection with multi-view data: A survey.
\newblock \emph{Information Fusion}, 50: 158--167.

\bibitem[{Zhang et~al.(2020{\natexlab{c}})Zhang, Wu, Cai, and
  Yu}]{zhang2020multi}
Zhang, Y.; Wu, J.; Cai, Z.; and Yu, P.~S. 2020{\natexlab{c}}.
\newblock Multi-view multi-label learning with sparse feature selection for
  image annotation.
\newblock \emph{IEEE Transactions on Multimedia}, 22(11): 2844--2857.

\end{thebibliography}


\begin{thebibliography}{}

\bibitem[\protect\citeauthoryear{Abelson \bgroup \em et al.\egroup
  }{1985}]{abelson-et-al:scheme}
Harold Abelson, Gerald~Jay Sussman, and Julie Sussman.
\newblock {\em Structure and Interpretation of Computer Programs}.
\newblock MIT Press, Cambridge, Massachusetts, 1985.

\bibitem[\protect\citeauthoryear{Baumgartner \bgroup \em et al.\egroup
  }{2001}]{bgf:Lixto}
Robert Baumgartner, Georg Gottlob, and Sergio Flesca.
\newblock Visual information extraction with {Lixto}.
\newblock In {\em Proceedings of the 27th International Conference on Very
  Large Databases}, pages 119--128, Rome, Italy, September 2001. Morgan
  Kaufmann.

\bibitem[\protect\citeauthoryear{Brachman and
  Schmolze}{1985}]{brachman-schmolze:kl-one}
Ronald~J. Brachman and James~G. Schmolze.
\newblock An overview of the {KL-ONE} knowledge representation system.
\newblock {\em Cognitive Science}, 9(2):171--216, April--June 1985.

\bibitem[\protect\citeauthoryear{Gottlob \bgroup \em et al.\egroup
  }{2002}]{gls:hypertrees}
Georg Gottlob, Nicola Leone, and Francesco Scarcello.
\newblock Hypertree decompositions and tractable queries.
\newblock {\em Journal of Computer and System Sciences}, 64(3):579--627, May
  2002.

\bibitem[\protect\citeauthoryear{Gottlob}{1992}]{gottlob:nonmon}
Georg Gottlob.
\newblock Complexity results for nonmonotonic logics.
\newblock {\em Journal of Logic and Computation}, 2(3):397--425, June 1992.

\bibitem[\protect\citeauthoryear{Levesque}{1984a}]{levesque:functional-foundations}
Hector~J. Levesque.
\newblock Foundations of a functional approach to knowledge representation.
\newblock {\em Artificial Intelligence}, 23(2):155--212, July 1984.

\bibitem[\protect\citeauthoryear{Levesque}{1984b}]{levesque:belief}
Hector~J. Levesque.
\newblock A logic of implicit and explicit belief.
\newblock In {\em Proceedings of the Fourth National Conference on Artificial
  Intelligence}, pages 198--202, Austin, Texas, August 1984. American
  Association for Artificial Intelligence.

\bibitem[\protect\citeauthoryear{Nebel}{2000}]{nebel:jair-2000}
Bernhard Nebel.
\newblock On the compilability and expressive power of propositional planning
  formalisms.
\newblock {\em Journal of Artificial Intelligence Research}, 12:271--315, 2000.

\end{thebibliography}

\end{document}